
\tolerance=10000
\input phyzzx

\REF\dwn{ B. de Wit and H. Nicolai,  {\it Phys. Lett.} {\bf 108 B} (1982) 285; 
B.
de Wit and H. Nicolai,  {\it Nucl. Phys.} {\bf B208} (1982) 323.}
\REF\nct{C.M. Hull, {\it Phys. Rev.} {\bf D30} (1984) 760; C.M. Hull,
{\it Phys. Lett.} {\bf 142B} (1984) 39; C.M. Hull, {\it Phys. Lett.} {\bf
148B} (1984) 297; C.M. Hull, {\it Physica} {\bf 15D} (1985) 230; Nucl.
Phys. {\bf B253} (1985) 650.}
\REF\nctt{ 
  C.M. Hull, {\it Class. Quant. Grav.} {\bf
2} (1985) 343.}
\REF\CJ{E. Cremmer and B. Julia, Phys. Lett. {\bf 80B} (1978) 48; Nucl. Phys.
{\bf B159} (1979) 141.}
\REF\frefo{ F. Cordaro, P. Fr\'e, L. Gualtieri, P. Termonia and M. Trigiante,
{\it $N=8$ gaugings revisited: an exhaustive classification}, Nucl. Phys. {\bf
B532} (1998) 245.}
\REF\CW{C.M. Hull and N. P. Warner, Class. Quant. Grav. {\bf 5} (1988)
1517.}
\REF\GH{
G.~W.~Gibbons and C.~M.~Hull,
{\it de Sitter space from warped supergravity solutions,}
hep-th/0111072.}
\REF\Cremmer{E. Cremmer, in {\it Supergravity and Superspace},
S.W.
Hawking and
M. Ro\v cek, C.U.P.
Cambridge,  1981.}
\REF\CT{
E.~Cremmer, B.~Julia, H.~Lu and C.~N.~Pope,
Nucl.\ Phys.\ B {\bf 523}, 73 (1998)
[hep-th/9710119].
}
\REF\Fer{
L.~Andrianopoli, R.~D'Auria, S.~Ferrara and M.~A.~Lledo,
hep-th/0203206.}
\REF\GRW{ M. G\"unaydin, L.J. Romans and N.P. Warner,
Phys. Lett. {\bf 164B}, (1985) 309.}
\REF\gunwar{ M. G\"unaydin, L.J. Romans and N.P. Warner, {\it Gauged $N=8$
Supergravity in Five Dimensions}, Phys. Lett. {\bf 154B},  4 (1985) 268; {\it
Compact and Non--Compact Gauged Supergravity Theories in Five Dimensions}, 
Nucl.
Phys. {\bf B272} (1986) 598.}
\REF\PPV{ M. Pernici, K. Pilch and P. van Nieuwenhuizen, {\it Gauged
$N=8~D=5$ Supergravity} Nucl. Phys. {\bf B259} (1985) 460.}
\REF\HT{C.M. Hull and P.K. Townsend, Nucl. Phys. {\bf B438} (1995)
109;  hep-th/9410167.}
\REF\topo{
B.~de Wit, C.~M.~Hull and M.~Rocek,
Phys.\ Lett.\ B {\bf 184}, 233 (1987).}
\REF\schur{N.P. Warner,  Phys. Lett. {\bf 128B},   (1983) 169;  Nucl. Phys. 
{\bf B231} (1984)
250.}
\REF\HWpot{C.~M.~Hull and N.~P.~Warner,
Nucl.\ Phys.\ B {\bf 253}, 650 (1985)
and 
Nucl.\ Phys.\ B {\bf 253}, 675 (1985).}
\REF\DS{C.M. Hull, {\it 
De Sitter space in supergravity and M theory},
JHEP {\bf 0111}, 012 (2001),
hep-th/0109213.}
\REF\wall{C.M. Hull, {\it 
 Domain wall and de Sitter solutions of gauged supergravity},
JHEP {\bf 0111}, 061 (2001), hep-th/0110048.}
\REF\quin{
P.~K.~Townsend,
JHEP {\bf 0111}, 042 (2001)
hep-th/0110072.}



\font\mybb=msbm10 at 12pt
\def\bbbb#1{\hbox{\mybb#1}}
\def\Z {\bbbb{Z}}
\def\R {\bbbb{R}}
\def\C {\bbbb{C}}
\def\lsemidir{\mathbin{\hbox{\hskip2pt\vrule height 5.7pt depth -.3pt width
.25pt\hskip-2pt$\times$}}}

\def\sd{{\lsemidir}} 
\def \aa {\alpha}
\def \bb {\beta}
\def \gg {\gamma}
\def \dd {\delta}
\def \ee {\epsilon}

\def \ll {\lambda}
\def \mm {\mu}
\def \nn {\nu}

\def \ss {\sigma}

\def \lll {\Lambda}

\def \sss {\Sigma}

\def \www{\Omega}

\def \ti {\tilde}

\def \2 {{1 \over 2}}
\def \3 {{1 \over 3}}
\def \4 {{1 \over 4}}
\def \5 {{1 \over 5}}
\def \6 {{1 \over 6}}
\def \7 {{1 \over 7}}
\def \8 {{1 \over 8}}
\def \9 {{1 \over 9}}
\def \0 { \infty}

\def\++ {{(+)}}
\def \- {{(-)}}
\def\+-{{(\pm)}}

\def\ek {\eqn\abc$$}

\def \pa {\partial}

\def \qq {\qquad}


 \def\unit{\hbox to 3.3pt{\hskip1.3pt \vrule height 7pt width .4pt \hskip.7pt
\vrule height 7.85pt width .4pt \kern-2.4pt
\hrulefill \kern-3pt
\raise 4pt\hbox{\char'40}}}

\def\nup#1({Nucl.\ Phys.\  {\bf B#1}\ (}

\def\cV {{\cal{V}}}
\def\cB {{\cal{B}}}
\def\cD {{\cal{D}}}

\def\tr  {{\rm tr}}


%
%
%

%
\newhelp\stablestylehelp{You must choose a style between 0 and 3.}%
\newhelp\stablelinehelp{You should not use special hrules when stretching%
a table.}%
\newhelp\stablesmultiplehelp{You have tried to place an S-Table %
inside another%
S-Table.  I would recommend not going on.}%
%
%
\newdimen\stablesthinline
\stablesthinline=0.4pt
\newdimen\stablesthickline
\stablesthickline=1pt
%
%
\newif\ifstablesborderthin
\stablesborderthinfalse
\newif\ifstablesinternalthin
\stablesinternalthintrue
\newif\ifstablesomit
\newif\ifstablemode
\newif\ifstablesright
\stablesrightfalse
%
%
\newdimen\stablesbaselineskip
\newdimen\stableslineskip
\newdimen\stableslineskiplimit
%
%
\newcount\stablesmode
\newcount\stableslines
\newcount\stablestemp
\stablestemp=3
\newcount\stablescount
\stablescount=0
\newcount\stableslinet
\stableslinet=0
%
%
%
\newcount\stablestyle
\stablestyle=0
%
%
\def\stablesleft{\quad\hfil}%
\def\stablesright{\hfil\quad}%
%
%
\catcode`\|=\active%
%
%
\newcount\stablestrutsize
\newbox\stablestrutbox
\setbox\stablestrutbox=\hbox{\vrule height10pt depth5pt width0pt}
\def\stablestrut{\relax\ifmmode%
                         \copy\stablestrutbox%
                       \else%
                         \unhcopy\stablestrutbox%
                       \fi}%
%
%
\newdimen\stablesborderwidth
\newdimen\stablesinternalwidth
\newdimen\stablesdummy
\newcount\stablesdummyc
\newif\ifstablesin
\stablesinfalse
%
%
\def\begintable{\stablestart%
  \stablemodetrue%
  \stablesadj%
  \halign%
  \stablesdef}%
\def\stablesadj{%
  \ifcase\stablestyle%
    \hbox to \hsize\bgroup\hss\vbox\bgroup%
  \or%
    \hbox to \hsize\bgroup\vbox\bgroup%
  \or%
    \hbox to \hsize\bgroup\hss\vbox\bgroup%
  \or%
    \hbox\bgroup\vbox\bgroup%
  \else%
    \errhelp=\stablestylehelp%
    \errmessage{Invalid style selected, using default}%
    \hbox to \hsize\bgroup\hss\vbox\bgroup%
  \fi}%
\def\stablesend{\egroup%
  \ifcase\stablestyle%
    \hss\egroup%
  \or%
    \hss\egroup%
  \or%
    \egroup%
  \or%
    \egroup%
  \else%
    \hss\egroup%
  \fi}%
\def\stablestart{%
  \ifstablesin%
    \errhelp=\stablesmultiplehelp%
    \errmessage{An S-Table cannot be placed within an S-Table!}%
  \fi
  \global\stablesintrue%
  \global\advance\stablescount by 1%
  \message{<S-Tables Generating Table \number\stablescount}%
  \begingroup%
  \stablestrutsize=\ht\stablestrutbox%
  \advance\stablestrutsize by \dp\stablestrutbox%
  \ifstablesborderthin%
    \stablesborderwidth=\stablesthinline%
  \else%
    \stablesborderwidth=\stablesthickline%
  \fi%
  \ifstablesinternalthin%
    \stablesinternalwidth=\stablesthinline%
  \else%
    \stablesinternalwidth=\stablesthickline%
  \fi%
  \tabskip=0pt%
  \stablesbaselineskip=\baselineskip%
  \stableslineskip=\lineskip%
  \stableslineskiplimit=\lineskiplimit%
  \offinterlineskip%
  \def\borderrule{\vrule width \stablesborderwidth}%
  \def\internalrule{\vrule width \stablesinternalwidth}%
  \def\thinline{\noalign{\hrule height \stablesthinline}}%
  \def\thickline{\noalign{\hrule height \stablesthickline}}%
  \def\trule{\omit\leaders\hrule height \stablesthinline\hfill}%
  \def\ttrule{\omit\leaders\hrule height \stablesthickline\hfill}%
  \def\tttrule##1{\omit\leaders\hrule height ##1\hfill}%
  \def\stablesel{&\omit\global\stablesmode=0%
    \global\advance\stableslines by 1\borderrule\hfil\cr}%
  \def\el{\stablesel&}%
  \def\elt{\stablesel\thinline&}%
  \def\eltt{\stablesel\thickline&}%
  \def\elttt##1{\stablesel\noalign{\hrule height ##1}&}%
  \def\elspec{&\omit\hfil\borderrule\cr\omit\borderrule&%
              \ifstablemode%
              \else%
                \errhelp=\stablelinehelp%
                \errmessage{Special ruling will not display properly}%
              \fi}%
  \def\stmultispan##1{\mscount=##1 \loop\ifnum\mscount>3 \stspan\repeat}%
  \def\stspan{\span\omit \advance\mscount by -1}%
  \def\multicolumn##1{\omit\multiply\stablestemp by ##1%
     \stmultispan{\stablestemp}%
     \advance\stablesmode by ##1%
     \advance\stablesmode by -1%
     \stablestemp=3}%
  \def\multirow##1{\stablesdummyc=##1\parindent=0pt\setbox0\hbox\bgroup%
    \aftergroup\emultirow\let\temp=}
  \def\emultirow{\setbox1\vbox to\stablesdummyc\stablestrutsize%
    {\hsize\wd0\vfil\box0\vfil}%
    \ht1=\ht\stablestrutbox%
    \dp1=\dp\stablestrutbox%
    \box1}%
  \def\stpar##1{\vtop\bgroup\hsize ##1%
     \baselineskip=\stablesbaselineskip%
     \lineskip=\stableslineskip%

\lineskiplimit=\stableslineskiplimit\bgroup\aftergroup\estpar\let\temp=}%
  \def\estpar{\vskip 6pt\egroup}%
  \def\stparrow##1##2{\stablesdummy=##2%
     \setbox0=\vtop to ##1\stablestrutsize\bgroup%
     \hsize\stablesdummy%
     \baselineskip=\stablesbaselineskip%
     \lineskip=\stableslineskip%
     \lineskiplimit=\stableslineskiplimit%
     \bgroup\vfil\aftergroup\estparrow%
     \let\temp=}%
  \def\estparrow{\vfil\egroup%
     \ht0=\ht\stablestrutbox%
     \dp0=\dp\stablestrutbox%
     \wd0=\stablesdummy%
     \box0}%
  \def|{\global\advance\stablesmode by 1&&&}%
  \def\|{\global\advance\stablesmode by 1&\omit\vrule width 0pt%
         \hfil&&}%
  \def\vt{\global\advance\stablesmode by 1&\omit\vrule width
\stablesthinline%
          \hfil&&}%
  \def\vtt{\global\advance\stablesmode by 1&\omit\vrule width
\stablesthickline%
          \hfil&&}%
  \def\vttt##1{\global\advance\stablesmode by 1&\omit\vrule width ##1%
          \hfil&&}%
  \def\vtr{\global\advance\stablesmode by 1&\omit\hfil\vrule width%
           \stablesthinline&&}%
  \def\vttr{\global\advance\stablesmode by 1&\omit\hfil\vrule width%
            \stablesthickline&&}%
\def\vtttr##1{\global\advance\stablesmode by 1&\omit\hfil\vrule width ##1&&}%
\stableslines=0%
\stablesomitfalse}%
\def\stablesdef{\bgroup\stablestrut\borderrule##\tabskip=0pt plus 1fil%
  &\stablesleft##\stablesright%
  &##\ifstablesright\hfill\fi\internalrule\ifstablesright\else\hfill\fi%
  \tabskip 0pt&&##\hfil\tabskip=0pt plus 1fil%
  &\stablesleft##\stablesright%
  &##\ifstablesright\hfill\fi\internalrule\ifstablesright\else\hfill\fi%
  \tabskip=0pt\cr%
  \ifstablesborderthin%
    \thinline%
  \else%
    \thickline%
  \fi&%
}%
\def\endtable{\advance\stableslines by 1\advance\stablesmode by 1%
   \message{- Rows: \number\stableslines, Columns:
\number\stablesmode>}%
   \stablesel%
   \ifstablesborderthin%
     \thinline%
   \else%
     \thickline%
   \fi%
   \egroup\stablesend%
\endgroup%
\global\stablesinfalse}
%
%

\Pubnum{ \vbox{  \hbox {QMUL-PH-02-07}  
\hbox{hep-th/0204156} 
}}
\pubtype{}
\date{April 2002}

\titlepage

\title {\bf  New Gauged N=8, D=4 Supergravities  }

\author{C.M. Hull}
\address{Physics Department,
\break
Queen Mary, University of London,
\break
Mile End Road, London E1 4NS, U.K.}
\andaddress{Newton Institute, 20 Clarkson Road, Cambridge CB3 0EH, U.K.}


\abstract { New gaugings of four dimensional $N=8$ supergravity are
constructed, including one which has a Minkowski space vacuum that preserves
$N=2$ supersymmetry and in which the gauge group is broken  to $SU(3)\times
U(1)^2$. Previous gaugings used the form of the ungauged action which is
invariant under a rigid $SL(8,\R)$ symmetry and promoted a 28-dimensional
subgroup ($SO(8),SO(p,8-p)$ or the non-semi-simple contraction
$CSO(p,q,8-p-q)$) to a  local gauge group. Here, a dual form of the ungauged
action is used which is invariant under $SU^*(8)$ instead of
$SL(8,\R)$ and new theories are obtained by gauging 28-dimensional 
subgroups of $SU^*(8)$. The gauge groups are non-semi-simple 
and are different real forms of the $CSO(2p,8-2p)$ groups, denoted
$CSO^*(2p,8-2p)$, and the new theories   have a rigid $SU(2)$ symmetry.
The  five dimensional gauged $N=8$ supergravities are dimensionally reduced to
$D=4$. The $D=5,SO(p,6-p)$ gauge theories reduce, after a duality 
transformation,
to the $D=4,CSO(p,6-p,2)$ gauging while the
$SO^*(6)$ gauge theory  reduces to the $D=4, CSO^*(6,2)$ gauge theory.
The new theories are related to the old ones via an analytic continuation.
The non-semi-simple gaugings can be dualised to forms
with different gauge groups.}

\endpage


\chapter{Introduction}

In addition to the $N=8,D=4$ gauged supergravity of [\dwn] with gauge group
$SO(8)$, there is a   class of   $N=8,D=4$ gauged supergravities
with non-compact gauge group [\nct,\nctt].
The ungauged Cremmer-Julia $N=8$ supergravity [\CJ] has an
$SL(8,\R)$ global symmetry
of the action, and in each of the  gaugings of [\nct,\nctt], 
 a 28-dimensional subgroup $K$ of the $SL(8,\R)$ global symmetry is
promoted to a local symmetry, using the 28 vector fields in the $N=8$
supergravity supermultiplet.
In [\frefo], an exhaustive classification of such gaugings was given and
it was argued that the  only possible gauge groups are the ones found in
[\dwn,\nct,\nctt].  It will be shown here that nonetheless there are further
gaugings of the
$N=8$ theory that fall outside this class, in which the gauge group is
not a subgroup of $SL(8,\R)$. The key point is that there are other dual
forms of the ungauged 
$N=8$ theory in which the global symmetry of the action is some group $L$
that is not contained in $SL(8,\R)$ and there is the possibility of 
gauging a 28-dimensional subgroup $K$ of $L$.

The ungauged $N=8$ supergravity in $D=4$ of [\CJ] has a global
$E_{7(7)}$ symmetry and a local $SU(8)$ symmetry.
The
$E_{7(7)}$ is a duality symmetry of the equations of motion and
the Cremmer-Julia action [\CJ] is invariant under a maximal  $SL(8,\R)$
subgroup of this. The bosonic sector consists of the graviton,  28
vector fields  transforming as a {\bf 28} of $SL(8,\R)$
and
70 scalars taking values in  the coset
$E_7/SU(8)$.
Gauging of this theory entails promoting a 28-dimensional subgroup $K$
of $SL(8,\R)$ to a local symmetry. 
In [\dwn], the gauging with $K=SO(8)$ was constructed, and in [\nct,\nctt]
gaugings were   constructed with non-compact gauge groups
$K=SO(p,8-p)$ or
certain non-semi-simple gauge groups
$CSO(p,q,r)$   
for all non-negative integers $p,q,r$ with
$p+q+r=8$,
and these are the only possible gaugings of subgroups of $SL(8,\R)$ [\frefo].  
When
$r=0$,
$CSO(p,q,0)=SO(p,q)$, while if $q=0$ the notation $CSO(p,0,r)=CSO(p,r)$
was used. The $CSO(p,q,r)$ group [\nctt] arises from a group contraction of
$SO(p+r,q)$.  Decomposing the
generators $L=\lll+\www+\sss$ of $SO(p+r,q)$ into  the generators $\lll$ 
of 
$SO(p,q)$,
the generators
$\www$ of $SO(r)$ and the remaining $r(p+q)$ generators $\sss$
and performing the rescaling 
$$L \to \lll+\xi \www+ \sqrt{\xi} \sss
\eqn\contr$$
and taking the contraction $\xi \to 0$ gives the $CSO(p,q,r)$ algebra.
In the limit, the 
generators of $CSO(p,q,r)$ then 
consist of the generators $\lll$ of $SO(p,q)$,   $r(r-1)/2$
abelian generators
$\www$, and $r(p+q)$ generators $\sss$ transforming
as $r$ copies of the vector representation of $SO(p,q)$.
The
commutation relations of
  $CSO(p,q,r)$ take the schematic form
$$ [\lll , \lll ] \sim \lll, \qq
[\lll, \sss ] \sim \sss, \qq [\sss , \sss ]\sim \www
\eqn\salg$$
with all other commutation relations vanishing.
The abelian generators $\www$ are central charges, appearing only on the
right hand side of $[\sss,\sss]$, so that the algebra is a central
extension of the semi-direct product
$SO(p,q)\sd  \R^{r(p+q)}$. 
There are other real forms of  $CSO(p,q,r)$ 
in which $U(1)^{r(r-1)/2}$
  is replaced by e.g. $U(1)^{r(r-1)/2-n}\times SO(1,1)^n$.
The $D=11$ origin of these theories
was found in [\CW] and analysed in [\GH].

However, there are other forms of the 
ungauged supergravity action related to the Cremmer-Julia theory by
duality transformations which have different global duality symmetry
groups
$G$ and  for each there will be a   subgroup $L\subset G$ which
is a symmetry of the action.
For example, the ungauged $N=8,D=5$ supergravity has a global $E_{6(6)}$
symmetry of the action [\Cremmer] and dimensional reduction gives a form
of the $N=8,D=4$ supergravity with a symmetry of the
action that is $L=E_{6(6)}\times
\R ^+$, which is not contained in $SL(8,\R)$.
Duality transformations take this to the  Cremmer-Julia form with
$G=E_{7(7)}$ and $L=SL(8,\R)$.
A dual form in which the symmetry of the action is $SU^*(8)$ will play an
important role here. Some other dual forms of the theory werere considered in
[\CT]. For any of these dual forms, there is the possibility of promoting a
subgroup $K$ of the symmetry $L$ of the action  to a local symmetry and seeking 
further modifications to the theory to preserve supersymmetry.
 If the gauge group $K$ is not a subgroup of $SL(8,\R)$,
such a gauging cannot be one of the ones previously found   and so must be new.
This point was also noted in [\Fer], where Scherk-Schwarz reductions of
$N=8,D=5$ gauged supergravity were shown to correspond to gaugings of
$N=8,D=4$ supergravity in which   flat subgroups of
$L=E_6\times \R^+$ are promoted to local symmetries.

The purpose here is to exhibit such   new gaugings.
It will be seen that there is a dual form of the $N=8$ action in which the
subgroup
$L$ of
$E_{7(7)}$ which is a symmetry of the action is $SU^*(8)$  instead of
$SL(8,\R)$, and in this formulation one can seek to gauge 28-dimensional
subgroups of
$SU^*(8)$. (Recall that $SU^*(2n)$ and $SO^*(2n)$ are certain non-compact
forms of $SU(2n)$ and $SO(2n)$, respectively; see section 3 for
definitions and discussion of these groups.)
 Such gauge theories will be constructed here with gauge groups
$CSO^*(2p,8-2p)$ for $p=1,2,3$. These gauge groups are non-semi-simple
contractions of
$ SO^*(8)$, defined (in section 3) in analogy with the contractions
$CSO(q,8-q)$ of $SO(8)$;
they are not  subgroups of $SL(8,\R)$.

The existence of one of these new gaugings was anticipated in [\GRW].
 Gauged $D=5,N=8$ theories were constructed with gauge group  
$SO(6) \sim SU(4)$ in [\gunwar,\PPV],  and with gauge groups $SO(5,1) \sim
SU^*(4)$,
$SO(4,2) \sim SU(2,2)$ and
$SO(3,3) \sim SL(4,\R)$ in [\gunwar].
 In [\GRW] a further gauging was found with
gauge group $SO^*(6) \sim SU(3,1)$.
 The $SU(3,1)$ gauging has a critical point with zero cosmological
constant that preserves $N=2$ supersymmetry and breaks the gauge group
down to $SU(3)\times U(1)$. It also has a global $SU(2)$ symmetry.
As discussed in [\GRW], this theory can be compactified to four dimensions to
give a theory
which also has a Minkowski space vacuum preserving 
$N=2$ supersymmetry. This theory will be constructed here and shown to
be a gauged $N=8$ supergravity.
It will be seen that reduction of the the $D=5$ theories with gauge groups
$SO(p,6-p)$ to $D=4$ and performing a duality transformation gives the
$D=4$ theories of [\nctt] with gauge groups
$CSO(p,6-p,2)$, which are non-semi-simple groups with a subgroup
$SO(p,6-p)\times SO(2)$. The commutation relations are of the form
\salg\ with $\lll$ generating $SO(p,6-p)$, a single abelian generator $\www$
and 12 generators $\sss$ in the $({\bf 6,2})$ representation
of $SO(p,6-p)\times SO(2)$. There is in addition a global
$SL(2,\R)$ symmetry, which is broken to $SL(2,\Z)$ in the quantum theory.
It will also be shown  that reducing the $SO^*(6)$ gauge theory
gives a new theory with
 gauge group 
$CSO^*(6,2)$, which is non-semi-simple with a   subgroup
$SU(3,1)\times U(1)$, and maximal compact subgroup
$SU(3)\times U(1)\times U(1)$. This unfortunately does not contain the
group $SU(3)\times SU(2)\times U(1)$ hoped for in [\GRW], but there is in
addition a global $SU(2)$ symmetry, broken to a discrete subgroup in the
quantum theory. The commutation relations are of the form
\salg\ with $\lll$ generating $SO^*(6)=SU(3,1)$, a single abelian generator
$\www$ and 12 generators $\sss$ in the $({\bf 6,2})$ representation
of $SO^*(6)\times SO(2)$.

The plan of the paper is as follows.
In section 2, the $CSO(p,q,r)$ gaugings of $N=8$ supergravity will be reviewed.
In section 3, the $SO^*(2n)$ and $SU^*(2n)$ groups will be 
reviewed and 
the duality symmetry of the $N=8$ theory discussed, and 
the form of the ungauged theory with action invariant under $SU^*(8)$ will be 
constructed.
In section 4, the theories in which  $CSO^*(2p,2q)$ subgroups of $SU^*(8)$ are
gauged are constructed for $p=1,2,3$ and their symmetries are discussed.
In section 5, the $N=8,D=5$ gauged supergravities are dimensionally reduced to
$D=4$ and the resulting $D=4$ gaugings identified, giving a
different derivation of the $CSO^*(6,2)$ theory.
In section 6, the action of dualities on these theories is considered.
For the non-semi-simple gaugings, the dualisation of some of the gauge fields 
is
possible, changing the gauge group.
Then U-duality transformations are considered and suitable limits of these are
shown to construct the new gaugings from old ones.
In section 7,  the scalar potentials are analysed and some of the physical
properties discussed.

\chapter{The $CSO(p,q,r)$ {} $D=4$ Gauged Supergravities}

The $CSO(p,q,r)$ gaugings arise from promoting a 
 28-dimensional subgroup $K$
of $SL(8,\R)$ to a local symmetry.
The 28 vector fields
become the gauge bosons, so that it is necessary that the subgroup $K$ is
chosen so that the  {\bf 28} of $SL(8,\R)$ becomes the adjoint of $K$.
Then supersymmetry requires the addition of terms depending on the
coupling constant
$g$   to the action and supersymmetry transformation rules, including a
scalar potential proportional to $g^2$.
In [\dwn], the gauging with $K=SO(8)$ was constructed, and in [\nct,\nctt]
gaugings were   constructed with non-compact gauge groups
$K=SO(p,8-p)$ or
the non-semi-simple gauge groups
$CSO(p,q,r)$   
for all non-negative integers $p,q,r$ with
$p+q+r=8$.
The group
 $CSO(p,q,r)$ is 
the group contraction of $SO(p+r,q)$
 preserving a symmetric metric with $p$ positive eigenvalues, $q$
negative ones and $r$ zero eigenvalues. Then  $CSO(p,q,0)=SO(p,q)$ and
$CSO(p,q,1)=ISO(p,q)$. 
The Lie algebra of $CSO(p,q,r)$ is [\nctt]
$$
[ L_{ab}, L_{cd} ] = L_{ad} \eta_{bc} -L_{ac} \eta_{bd} -
L_{bd} \eta_{ac} +L_{bc} \eta_{ad}
\eqn\alg$$
where
$$ \eta_{ab} =\pmatrix{
{\bf 1}_{p \times p} & 0 &0 \cr
0 & -{\bf 1}_{q \times q} &0\cr 0& 0& 0 _{r\times r}
}\eqn\abc$$     
  $a, b = 1, \cdots, 8$ and $L_{ab}=-L_{ba}$. 
Note that despite the non-compact
gauge groups, these are unitary theories, as the vector kinetic term is
not the minimal term
 constructed with the indefinite Cartan-Killing metric,  but is
constructed  
with a positive definite scalar-dependent matrix. The $CSO(p,q,r)$
gauging and the $CSO(q,p,r)$
gauging are equivalent.

The 70 scalars  
  parameterise the coset $E_{7(7)}/SU(8)$, while the
28 vector fields $A_\mm $ transform as a {\bf 28} of the
subgroup $SL(8,\R)\subset E_{7(7)}$.
The 28 field strengths $F$ satisfy the Bianchi identitites $dF=0$, and it
is useful to define 28 dual 2-forms $G$, transforming in the
${\bf 28 '}$ of  $SL(8,\R)$,
so that the
field equations are of the form $dG=0$. In the linearised theory, $G\sim
*F$, but in the full interacting theory there are 
field-dependent modifications to this [\CJ].
Then the 28+28 field strengths $(F,G)$ combine into
a 56-vector transforming as the {\bf 56} representation of
$E_{7(7)}$.

Different bases for 
$E_{7(7)}$ are useful for different purposes.
In [\dwn], a basis in which the
$SU(8)$ subgroup of $E_{7(7)}$ is manifest was used, with the
 {\bf 56}   of
$E_{7(7)}$ decomposing into a ${\bf 28} + {\bf   \overline{28}}$ of $SU(8)$.
In this basis, the formulae for the non-compact gaugings are rather
complicated, and are dramatically simplified by going to
the basis in which
the subgroup $SL(8,\R)\subset E_{7(7)}$ is manifest, with
${\bf 56} \to {\bf 28} + {\bf  28'}$.
Details of how to transform between these bases are given in [\CJ,\nct].
Let upper indices $a,b=1,...,8$ label the {\bf 8} of $SL(8,\R)$ and
lower indices $a,b$ label
the contragredient ${\bf 8'}$.
Then the field strengths in the {\bf 28}  are
$F^{ab}_{\mm\nn} =-F^{ba}_{\mm\nn}
$, while the dual field strength 2-forms  are $G_{ab}=-
G_{ba}$.
As in [\CJ], we introduce
indices $i,j=1,...,8$ in the {\bf 8} of $SU(8)$, which are raised and
lowered by complex conjugation.
The scalar fields can be represented by
a $56\times 56$ matrix $\cV$.
$$
\cV=
\pmatrix{u_{ij}^{\;\;\;ab}  &  v_{ijcd} \cr
\overline{v}^{klab} & \overline{u}^{kl}_{\;\;\;cd}}
\eqn\viss$$
transforming under
a rigid $E_7$ transformation represented by a $56\times 56$ matrix $E$
and a
local $SU(8)$ transformation $U(x)$
represented by a $56\times 56$ matrix in the $SU(8)$ subgroup of $E_7$ as
$$
\cV\to U(x)\cV E^{-1}
\eqn\abc$$
The scalar kinetic term
in the ungauged theory can be written as
$$
\int d^4x \sqrt {g} \, {\rm tr}  (D_\mu \cV \cV^{-1} D^\mu \cV \cV^{-1} )
\eqn\abc$$
where
$D_\mu$ is an $SU(8)$ covariant derivative involving an $SU(8)$
connection
$\cB _{\mu i} {}^j
$. The $SU(8)$ connection appears algebraically in the action and its
field equation   determines $\cB _{\mu i} {}^j
$ in terms of the other fields.

Minimal coupling for the group $K=CSO(p,q,r)$  
uses the fact that the {\bf 28} of $SL(8,\R)$ is the adjoint of $K\subset 
SL(8,\R)$
and consists
of introducing the non-abelian field strengths
$$F_{\mu \nu}^{ab} = \pa_{\mu} A_{\nu}^{ab} - \pa_{\nu} A_{\mu}^{ab}-2 g
A_{[\mu}^{ca} A_{\nu]}^{bd}\eta _{cd}
\eqn\fis$$
with gauge coupling $g$
and replacing the $SU(8)$ covariant derivative $D_\mm$ with 
the  $SU(8)\times K$ covariant derivative $\cD _\mm$, so that for example
$$
\cD_\mm u_{ij}{}^{ab}
=\pa _\mm u_{ij}{}^{ab} + \cB_\mm {}^k{}_{[i}
 u_{j]k}{}^{ab}-2g
A_\mm
 ^{c[a}
u_{ij}{}^{b]d}
\eta _{cd}
\eqn\minu$$

These minimal couplings break supersymmetry, but supersymmetry can be
restored by $g$-dependent modifications of    the action and
supersymmetry transformations. These involve scalar-dependent $SU(8)$
tensors
$A_1^{\;\;ij} ,A_{2l}^{\;\;\;ijk}$
given in terms of
  the
   $T$-tensor
$$T_i^{\;jkl}  = {v}^{kl cd} 
\eta_{ad}
 \left( u_{im}^{\;\;\;ab} 
{u}^{jm}_{\;\;\;bc}-v_{im ab} {v}^{jm bc} \right)
\eqn\tis$$
by
$$
A_1^{\;\;ij}  = 
-{4\over 21} T_{m}^{\;\;ijm}, \;\;\; A_{2l}^{\;\;\;ijk}=-{4\over 3}
T_{l}^{\;[ijk]},
\eqn\ais$$
The scalar-dependent T-tensor encodes all the $g$-dependent   corrections to 
the
theory and checking supersymmetry entails showing that the T-tensor satisfies 
a number of non-trivial identities.
While the T-tensor for the non-compact gaugings takes the relatively simple 
form 
\tis\
in this basis for $E_7$ in which $SL(8,\R)$ is manifest, transforming to a 
basis 
 in which $SU(8) $ is manifest results in an
expression for the T-tensor which is considerably more complicated (see 
equation 
(24)
of [\nctt]).
 The extra terms in the action consist of the scalar potential  
$$
V= -g^2 \left( {3\over 4} \left\vert A_1^{\;ij}  \right\vert ^2-
{1\over24} 
\left\vert  A^{\;\;i}_{2\;\;jkl} \right\vert ^2 \right), 
\eqn\vis$$
and the fermion bilinear terms
$$\eqalign{
L_g&= \sqrt 2 g A_{1\, ij} \bar \psi_{\mu}^i
\gg ^{\mm \nn } \psi_{\nu}^j + \6 g A_{2l}^{\;\;\;ijk}  \bar \psi_{\mu}^l 
\gg^\mm
\chi _{ijk} 
\cr &
+{1\over 144} \sqrt 2 g \ee ^{ijkpqrlm} A_2{}^n {}_{pqr}
\bar\chi _{ijk}  \chi _{lmn}  +h.c.
\cr}
\eqn\ferms$$
while the modifications to the supersymmetry transformations are
$$\dd_g \psi_{\mu}^i  =  
  - \sqrt{2} g A_1^{\;ij} \gg_{\mu} \ee_j, 
\qq
\dd_g  \chi^{ijk}   =  - 2 g 
A_{2l}^{\;\;\;ijk} 
\ee^l.
\eqn\susyv$$
The conventions are as   in [\dwn,\nct], so that fermions with upper indices 
are
right-handed and ones with lower  indices are left-handed.
 Note that the $T$-tensor is invariant under
$K\times SL(r,\R)$ and transforms reducibly under $SU(8)$,
decomposing into 
the tensor $A_1$ transforming as 
  a {\bf 36}
and
the tensor $A_2$ transforming as 
  a {\bf  420}.

For the $CSO(p,q,r)$ gaugings with $r>0$, it is useful to decompose
the $SL(8,\R)$ indices $a,b=1,...,8$ 
into $SL(p+q,\R)$ indices $I,J=1,...,p+q$
and
 $SL(r,\R)$ indices $\aa,\bb=p+q+1,...,8$, so that $a\to (I,\aa)$ and
$$ \eta_{ab} =\pmatrix{
\eta _{IJ} &0 \cr   0& 0 _{r\times r}
}\eqn\etiss$$  
where $\eta _{IJ}=diag({\bf 1}_{p \times p},- {\bf 1}_{q \times q} )$ is
the  invariant metric of $SO(p,q)$.
Then $A^{ab}\to (A^{IJ}, A^{I\aa}=-A^{\aa I}
,A^{\aa\bb})$ and
$$\eqalign{
F_{\mu \nu}^{IJ} &= \pa_{\mu} A_{\nu}^{IJ} - \pa_{\nu}
A_{\mu}^{IJ}-2 g A_{[\mu}^{KI} A_{\nu]}^{JL}\eta _{KL}
\cr
F_{\mu \nu}^{I\aa} &= \pa_{\mu} A_{\nu}^{I\aa} - \pa_{\nu}
A_{\mu}^{I\aa}-2 g A_{[\mu}^{KI} A_{\nu]}^{\aa L}\eta _{KL}
\cr
F_{\mu \nu}^{\aa \bb} &= \pa_{\mu} A_{\nu}^{\aa \bb} - \pa_{\nu}
A_{\mu}^{\aa \bb}-2 g A_{[\mu}^{K\aa} A_{\nu]}^{\bb L}\eta _{KL}
\cr}
\eqn\fiss$$
Similarly, $u_{ij}{}^{ab} \to
(u_{ij}{}^{IJ},u_{ij}{}^{I\aa},u_{ij}{}^{\aa\bb})$ etc, and the T-tensor
becomes
$$T_i^{\;jkl}  = {v}^{kl cJ} 
\eta_{IJ}
 \left( u_{im}^{\;\;\;Ib} 
{u}^{jm}_{\;\;\;bc}-v_{im Ib} {v}^{jm bc} \right)
\eqn\tiss$$
and the invariance of this under $SO(p,q)\times SL(r,\R)$ is manifest.
The covariant derivative can also be written in this basis, so that e.g. \minu\
becomes
$$
\eqalign{
\cD_\mm u_{ij}{}^{IJ}
&=\pa _\mm u_{ij}{}^{IJ} + \cB_\mm {}^k{}_{[i}
 u_{j]k}{}^{IJ}-2g
A_\mm
 ^{K[I}
u_{ij}{}^{J]L}
\eta _{KL}
\cr
\cD_\mm u_{ij}{}^{I\aa }
&=\pa _\mm u_{ij}{}^{I\aa } + \cB_\mm {}^k{}_{[i}
 u_{j]k}{}^{I\aa}-2g
A_\mm
 ^{K[I}
u_{ij}{}^{\aa]L}
\eta _{KL}
\cr
\cD_\mm u_{ij}{}^{\aa \bb}
&=\pa _\mm u_{ij}{}^{\aa \bb} + \cB_\mm {}^k{}_{[i}
 u_{j]k}{}^{\aa \bb}-2g
A_\mm
 ^{K[\aa }
u_{ij}{}^{\bb]L}
\eta _{KL}
\cr}
\eqn\duis$$

Note that $A^{\aa\bb}$ only appears through its curl
$dA^{\aa\bb}$ in \fiss,\duis.

\chapter{Group Structure and Bases}

The new gaugings involve the groups $SO^*(2n)$ and $SU^*(2n)$, and before 
proceeding
it will be useful to recall their definitions.
The group $SO^*(2n)$ is the non-compact form of $SO(2n)$ whose maximal compact
subgroup is $U(n)$, while $SU^*(2n)$ is the non-compact form of $SU(2n)$ whose 
maximal
compact subgroup is $USp(2n)$.
Equivalently, these groups can be defined as subgroups of $SL(2n,\C)$.
Let $J$ be an antisymmetric $2n\times 2n$  matrix satisfying $J^2=-1$.
The subgroup  $SU^*(2n)$ of $SL(2n,\C)$ consists of those $2n\times 2n$
complex matrices 
$U$  in  $SL(2n,\C)$ satisfying
$$ UJ=JU^*
\eqn\uni$$
where $U^*$ is the complex conjugate of $U$.
The subgroup $SO^*(2n)$ of $SU^*(2n)$ consists of those matrices $U$ also 
satisfying
$$
UU^t=1
\eqn\orth$$
where $U^t$ is the transpose of $U$. Thus $SO^*(2n)$ preserves the metric 
$\dd_{ij}$
as well as the complex structure $J$, since
$$UJU^\dagger =J
\eqn\jpres$$
The subgroup of $SL(2n,\C)$ of matrices satisfying \orth\ (but not
necessarily 
\uni) is $SO(2n,\C)$ and so $SO^*(2n)$ is also the subgroup of
$SO(2n,\C)$ preserving a complex structure, \jpres.

Writing $U=e^\lll$ in terms of Lie algebra generators
$\lll ^a{}_b$ ($a,b=1,....,2n$), 
the generators of $SL(2n,\C)$ are general traceless complex matrices
$\lll ^a{}_b$, $\lll ^a{}_a=0$, while the generators of
$SU^*(2n)$ are those satisfying in addition
$$
\lll^a{}_c J^{cb}=J^{ac} \lll_c{}^b
\eqn\abc$$
where
$$
\lll _a {}^b =(\lll^a{}_b)^*\eqn\abc$$
and the matrix $J^{ab}$
is real and satisfies
$$J^{ab}=J_{ab} 
=-J^{ba},\qq J^{ab}J_{bc}=-\dd^a{}_c
\eqn\abc$$
Defining
$$\lll_{ab}=J_{ac}\lll _c {}^b, \qq
\lll^{ab}=(\lll_{ab})^*
\eqn\abc$$
this implies
$$\lll^{ab}=J^{ac}J^{bd}\lll_{cd}
\eqn\abc$$
The generators of $SO^*(2n)$ satisfy in addition
$$\dd _{ac}\lll ^c{} _b =-\dd _{bc}\lll ^c{} _a
\eqn\abc$$
Defining
$$ 
L_{ab}=\dd _{ac}\lll ^c{} _b, 
\eqn\abc$$
the subgroup
$SO(2n,\C)$ of $SL(2n,\C)$
is generated by complex antisymmetric matrices satisfying
$$L_{ab}=-L_{ba}
\eqn\abc$$
Then $SO^*(2n)$ is defined as the subgroup of $SO(2n,\C)$
for which the generators satisfy
the reality condition
$$L^{ab}=J^{ac}J^{bd} L_{cd}
\eqn\realll$$
where
$L^{ab}\equiv (L_{ab})^*$.
This is to be compared to the group $SO(2n)$, which is defined as the
subgroup satisfying the standard reality condition that the generators
are real,
$L_{ab}\equiv (L_{ab})^*$.

Some of the lower dimensional cases are (locally) equivalent to more
familiar groups:
$$ \eqalign{& SO^*(2) \sim SO(2), \qq SO^*(4) \sim SO(3)\times SO(2,1),
\qq SO^*(6) \sim SU(3,1), 
\cr &
SO^*(8) \sim SO(6,2),\qq SU^*(2) \sim SU(2),\qq  SU^*(4) \sim SO(5,1)
\cr}
\eqn\abc$$

The $N=8$ supermultiplet has 28 vector fields.
In the abelian theory, the 28 field strengths and their 28 duals combine into 
the
{\bf 56} of $E_{7(7)}$ and the $E_{7(7)}$ transformations include 
duality transformations mixing electric and magnetic fields.
In formulating the theory, one chooses 
  28 of the 56 field strengths $F^A$ to be  given in terms of   fundamental 
potentials
$F^B=dA^B$ ($A,B=1,....,28$), and the other 28 $G_A$ to be given by the 
variation 
of
the action $S$, 
$G_A \sim * \dd S/ \dd F^A$, so that $G_A=*F^A+....$ (where the dots
denote terms dependent on the scalars and other fields). The vector field 
Bianchi identitites  and
field equations are then
$$dF^A=0, \qq dG_A=0
\eqn\abc$$
and these 56 equations transform into each other  as the {\bf 56} of $E_7$. 
Different choices of which set of 28 field strengths out of the 56 are to be
fundamental, i.e. which 28 are to be given  in terms of fundamental potentials,
give different dual formulations of the theory, all classically equivalent.

It is often  useful to choose a basis in which a subgroup of
$E_{7(7)}$ is a manifest symmetry. In [\CJ], two bases were considered, one in
which
$SU(8)\subset  E_{7(7)} $ is manifest, and one in which 
$SL(8,\R)\subset  E_{7(7)} $ is manifest.
Under $SL(8,\R)$, the {\bf 56} of $E_7$ decomposes into a ${\bf 28} +  {\bf 
28'}$.
Choosing the field strengths $F^{ab}$ in the ${\bf 28}$ (with $a,b=1,..,8$ 
$SL(8,\R)$ indices and $F^{ab}=-F^{ba}$) as fundamental,
$F^{ab}=dA^{ab}$, those in the  ${\bf 28'}$ are then the dual field strengths
$G_{ab}$.
In this formulation, $SL(8,\R)$ is a symmetry of the action while $E_7$ is a 
symmetry
of the equations of motion. This is the basis best suited to gauging the
$CSO(p,q,r)$ subgroups of 
$SL(8,\R)$. 
The field strengths can be combined into 
28 complex  combinations
${\cal F}=F+iG=F+i*F +...$ satisfying a generalised self-duality constraint 
[\CJ]
and  these transform as a {\bf 28} under $SU(8)\subset E_7$.
One can change to a basis in which this $SU(8)$ is manifest
(the explicit formulae relating tensors with $SL(8,\R)$ indices to ones with
$SU(8)$ indices involve 
  $SO(8)$ gamma-matrices [\CJ]) and in this basis the $SO(8)$ gauging is
straightforward, but the expressions for the T-tensor etc in the non-compact 
gaugings
look much more complicated than in the $SL(8,\R)$ basis.
Note that 
the 28 complex field strengths ${\cal F}$ cannot be written in terms of 28 
complex
potentials, and there is no formulation in which this $SU(8)\subset E_7$
is a rigid symmetry of the action, as it necessarily involves duality
transformations. The $SU(8)$ subgroup of the rigid $E_7$ symmetry should not be
confused with the local $SU(8)$ symmetry of the theory.

One can instead focus on any of the maximal subgroups of $E_{7(7)}$,
which are 
$$\eqalign{&SU(8), \qq SL(8,\R), \qq SU(4,4), \qq SU^*(8), \qq SO(6,6)\times 
SL(2,\R), \cr &
SO^*(12)\times SU(2),
\qq E_{6(2)}\times U(1),\qq  E_{6(6)}\times SO(1,1)\cr}
$$
The basis in which $SO(6,6)\times SL(2,\R)$ is manifest is useful in 
considering 
compactification from the $D=10$ IIB theory, with the $SO(6,6)$ the T-duality 
group
and  $SL(2,\R)$ the duality symmetry of the $D=10$ IIB supergravity [\HT].
The basis in which $E_{6(6)}\times SO(1,1)$ is   manifest is useful in 
considering
compactification from $D=5,N=8$ supergravity. The ungauged $D=5$ theory has 
 an $E_{6(6)}$ symmetry of the action, and so compactification gives
a dual form of $D=4,N=8$ supergravity which has
a symmetry of the action $E_{6(6)}\times SO(1,1)$.

Consider now the $SU^*(8)$ basis.
Introducing indices $a',b'...=1,....,8$ in the fundamental representation of
$SU^*(8)$, 
antisymmetric tensors $T^{a'b'}=-T^{b'a'}$ lie in a
representation of 28 complex dimensions.
However, this is reducible and one can impose the reality condition
$$ T_{ a'b'} \equiv (T ^{a'b'})^*=   J_{a'c'}J_{b'd'}T^{b'd'}
\eqn\real$$
using the complex structure of $SU^*(8)$ to define 
the {\bf 28} representation of 28 real dimensions.
From \realll, the  {\bf 28} of $SU^*(8)$ becomes the adjoint of $SO^*(8)\subset
SU^*(8)$. In addition, there is a dual real 28-dimensional representation, the
$   {\bf 28 '}$, defined by antisymmetric tensors $S_{a'b'}=-S_{b'a'}$
satisfying a reality condition similar to
\real\ and transforming so
  that $S_{a'b'}T^{a'b'}$ is a singlet.
Under $SU^*(8)$, the {\bf 56} of 
$E_{7(7)}$ decomposes as
$${\bf 56}  \to  {\bf 28} +   {\bf 28 '}  
\eqn\abc$$
Then
the 56 field strengths
can be decomposed into   fundamental ones $F^{a'b'}=-F^{b'a'}$
in the ${\bf 28}$ of $SU^*(8)$, and the duals
$G_{a'b'}$ in the $   {\bf 28 '}  $.
The fundamental field strengths
 satisfy  the reality condition
$$ F_{ a'b'} \equiv (F ^{a'b'})^*=  J_{a'c'}J_{b'd'}F^{b'd'}
\eqn\freal$$
and
are given in terms of potentials $A^{a'b'}$ satisfying a similar reality
condition, so that in the ungauged theory
$F^{a'b'}= dA^{a'b'}$.
The theory can then be
 formulated   in 
terms of the
28  real potentials $A^{a'b'}_\mu$. It is straightforward to do this, 
giving a
dual form of the theory in which 
$SU^*(8)$ is a symmetry of the action, with $SU^*(8)$  acting directly on the 
28
 real vector potentials in the {\bf 28}.
The action can be written in a way that is formally identical to 
the $SL(8,\R)$-invariant  Cremmer-Julia action, but with  $SL(8,\R)$ indices 
$a,b,...$ replaced by $SU^*(8)$ indices 
$a',b',...$, and the reality conditions on the fields changed to 
$SU^*(8)$-invariant ones of the type discussed above. 
In the next section, the  gauging of 28-dimensional subgroups of this 
$SU^*(8)$ rigid
symmetry will  considered.

\chapter{The $CSO^*(2p,8-2p)$ Gaugings}

The key to constructing these new gaugings is the observation that
the structure of the ungauged theory in the $SL(8,\R)$ basis is formally 
identical
to that in the $SU^*(8)$ basis, but with indices 
$a,b,...$ replaced by $SU^*(8)$ indices 
$a',b',...$. Then the structure of the gaugings of subgroups of
$SL(8,\R)$ when written in the $SL(8,\R)$ basis is   identical to
the structure of the gaugings of subgroups of
  $SU^*(8)$ when written in the   $SU^*(8)$ basis. 
The formal equivalence means that the proofs of the T-tensor identities are
formally the same, and hence the proof of supersymmetry for the new gaugings
follows immediately.
 Despite appearances, this is
not a trivial rewriting and the fact that the new T-tensors are defined with
contractions in the $SU^*(8)$ basis completely changes the  dependence of the
potential and mass terms on the scalar fields. This is   the same argument that 
was
   used in $D=5$ in [\GRW], where it was found that
the gauging of $SO(6)$ in the $SL(6,\R)$ basis and the gauging of $SO^*(6)$ in 
the
$SU^*(6)$ basis were formally identical, allowing the
 $SO^*(6)$ gauge theory to be written down immediately.

The group $CSO(p,q)$ arises from a group contraction of
$SO(p+q)$ in which the generators $L$ of 
$SO(p+q)$ are decomposed into $SO(p)$ generators $\lll$, $SO(q)$ generators 
$\www$
and the remaining generators $\sss$, then scaled as in \contr, and finally the
contraction 
$\xi \to 0$ is taken.
The resulting group has a subgroup $SO(p)\times U(1)^{q(q-1)/2}$.
It will be useful to define the group $CSO^*(2p,2q)$   by  an analogous
contraction of $SO^*(2p+2q)$. (Recall that $SO^*(n)$ is only defined for $n$
even.) Decomposing the generators $L$ of 
$SO^*(2p+2q)$   into $SO^*(2p)$ generators $\lll$, $SO^*(2q)$
generators $\www$ and the remaining generators $\sss$, then scaling as in 
\contr, and
finally taking the contraction 
$\xi \to 0$ gives the   group $CSO^*(2p,2q)$, which is a different real
form of
$CSO(2p,2q)$. It has a subgroup $SO^*(2p)\times U(1)^{q^2}\times
SO(1,1)^{q(q-1)}$ and commutation relations of the form \salg.
Both $CSO^*(2p,2q)$ and
$CSO(2p,2q)$ preserve a metric
$$ \eta  =\pmatrix{
{\bf 1}_{2p \times 2p} & 0 \cr
0 & {\bf 0}_{2q \times 2q} 
}\eqn\abc$$

The gauging of $CSO(2p,8-2p)$ in the $SL(8,\R)$ basis is as in section 2,
using the 28 vector fields $A^{ab}$ in the {\bf 28} of $SL(8,\R)$. The 
non-abelian
field strength is \fis, the minimal couplngs are as in \minu\ and then the 
further
modifications to the action and supersymmetry transformations are given in 
terms
of the T-tensor \tis.
The gauging of $CSO^*(2p,8-2p)$ in the $SU^*(8)$ basis is formally identical,
 but with  $SL(8,\R)$ indices 
$a,b,...$ replaced by $SU^*(8)$ indices 
$a',b',...$.
The 28 vector field strengths
$F^{a'b'}$ satisfying  \freal\ are defined in terms of
28 vector potentials
$A^{a'b'} $ satisfying the reality condition
$$ A_{ a'b'} \equiv (A ^{a'b'})^*=   J_{a'c'}J_{b'd'}A^{b'd'} 
\eqn\abc$$
by
$$F_{\mu \nu}^{a'b'} = \pa_{\mu} A_{\nu}^{a'b'} - \pa_{\nu} A_{\mu}^{a'b'}-2 g
A_{[\mu}^{c'a'} A_{\nu]}^{b'd'}\eta _{c'd'}
\eqn\fisp$$
where $\eta _{c'd'}$ is the invariant metric $diag(1_{2p}, 0_{8-2p})$.
In this basis, the components of the 56-bein
{\cal V} have the same structure as in  \viss, but with primed indices
$a',b',...$ instead of unprimed ones. The ovariant derivative of e.g.
$u_{ij}{}^{a'b'}$
is
$$
\cD_\mm u_{ij}{}^{a'b'}
=\pa _\mm u_{ij}{}^{a'b'} + \cB_\mm {}^k{}_{[i}
 u_{j]k}{}^{a'b'}-2g
A_\mm
 ^{c'[a'}
u_{ij}{}^{b']d'}
\eta _{c'd'}
\eqn\minup$$
The new $A$-tensors are given by \ais\ in terms of the new T-tensor  
$$T_i^{\;jkl}  = {v}^{kl c'd'} 
\dd_{a'd'}
 \left( u_{im}^{\;\;\;a'b'} 
{u}^{jm}_{\;\;\;b'c'}-v_{im a'b'} {v}^{jm b'c'} \right)
\eqn\tisp$$
Then adding the potential \vis\ and the mass terms \ferms\  and modifying the
supersymmetry transformations by the terms \susyv\ gives a theory which is
invariant
  under the $N=8$ local supersymmetry; the proof is formally equivalent to that 
for
the
$CSO(2p,8-2p)$ gauging. 
The $CSO(2p,8-2p)$ gauging has a rigid $SL(8-2p,\R)$ 
symmetry while the $CSO^*(2p,8-2p)$ gauging has a rigid $SU^*(8-2p)$ 
symmetry.

For the $CSO(p,q,r)$ gaugings considered in section 2, it was useful to 
decompose
the $SL(8,\R)$ indices $a,b=1,...,8$ 
into $SL(p+q,\R)$ indices $I,J=1,...,p+q$
and
 $SL(r,\R)$ indices $\aa,\bb=p+q+1,...,8$.
Similarly, for the
$CSO^*(2p,2q)$ gauging  with $p+q=4$, it is useful to decompose the
the   $SU^*(8)$ indices $a',b'=1,...,8$ 
into $SU^*(2p)$ indices $I',J'=1,...,2p$
and $SU^*(2q)$ 
  indices $\aa ',\bb '= 1,...,2q$, so that $a' \to (I',\aa ')$ and
$$ \eta_{a'b'} =\pmatrix{
\dd _{I'J'} &0 \cr   0& 0 _{2q \times 2q}
}\eqn\etxiss$$  
Then
$A^{a'b'}\to (A^{I'J'}, A^{I'\aa '} 
,A^{\aa '\bb '})$
with field strengths and covariant derivatives given by
\fiss,\duis, but with 
indices $I,\aa$ replaced by $I',\aa '$ and $\eta _{KL}$ replaced by
$\dd_{K'L'}$, while the T-tensor becomes
$$T_i^{\;jkl}  = {v}^{kl \aa 'J'} 
\eta_{I'J'}
 \left( u_{im}^{\;\;\;I'\bb '} 
{u}^{jm}_{\;\;\;\bb '\aa '}-v_{im I'\bb '} {v}^{jm \bb '\aa '} \right)
\eqn\tisst$$

This then gives a family of gaugings with local symmetry group $K=CSO^*(2p,2q)$
where $p+q=4$, and in addition a global symmetry $SU^*(2q)$.
The  gauge groups have a subgroup
$$ SO^*(2p)\times U(1)^{q^2}\times SO(1,1)^{q(q-1)}
\eqn\abc$$
and the maximal compact subgroup is
$$\ti K = SU(p)\times U(1)^{q^2+1} 
\eqn\abc$$
As usual, 
any solution will spontaneously break the
gauge symmetry to  a compact subgroup, contained in $\ti K$.

The case $q=0$ is the gauging of $SO^*(8)$. However, $SO^*(8)=SO(6,2)$
and is a common subgroup of both $SL(8,\R)$ and of $SU^*(8)$, so that
  gauging $SO^*(8)\subset SU^*(8)$  is equivalent to the
$SO(6,2)$ gauging of [\nct].
The $CSO^*(6,2)$ group has
a  maximal subgroup $SO^*(6)\times U(1)$ where
 $SO^*(6)\sim SU(3,1)$ with maximal compact subgroup $\ti K =
SU(3)\times U(1)^{2} $ and the theory has a global symmetry
$SU(2)$. This will be discussed further in later sections, where it will be
seen that there is an $N=2$ supersymmetric Minkowski space solution with 
gauge symmetry broken down to the maximal compact subgroup
$SU(3)\times U(1)^{2} $.

The $CSO^*(4,4)$ group has
a maximal subgroup 
$SO^*(4)\times U(1)^4\times SO(1,1)^2$ with
$SO^*(4)\sim SU(2)\times SU(1,1)$ and has  maximal compact
subgroup
$\ti K = SU(2)\times U(1)^{5} $. The theory has a global symmetry
$SU^*(4)\sim SO(5,1)$.
There is then a family of related gaugings with the following gauge groups $K$
 with maximal  subgroups $H\times C$ and global symmetry $U$:

\vskip 0.5cm
{\vbox{
\begintable
   $K$| $C$|$H$ | $U$  \elt 
    $CSO(4,0,4)$  |  $ U(1)^6$ | $SO(4)\sim SU(2)\times SU(2)$    |    
$SL(4,\R)
\sim SO(3,3)$ 
   \elt
    $CSO(3,1,4)$  | $ U(1)^6$ |  $SO(3,1)\sim SL(2,\C)$     |    
$SL(4,\R)
\sim SO(3,3$ 
   \elt
    $CSO(2,2,4)$  | $
 U(1)^6 $|  $SO(2,2)\sim SU(1,1)\times
SU(1,1)$     |    
$SL(4,\R)
\sim SO(3,3$ 
   \elt
    $CSO^*(4,4)$  |  $
 U(1)^4\times SO(1,1)^2$| $SO^*(4)\sim
SU(2)\times SU(1,1)$     |     $ SU^*(4)
 \sim SO^*(6)
$ 
     \endtable
}}
\vskip .5cm

The $CSO^*(2,6)$ gauging has
a subgroup 
$ U(1)^{10}\times SO(1,1)^{6}$ (recall $SO^*(2)=U(1)$), maximal compact 
subgroup
$\ti K =   U(1)^{10} $ and global symmetry
$SU^*(6)$.
This is distinct from the
 $CSO(2,6)$ gauging which has
a   maximal compact subgroup
$    U(1)^{16} $ and global symmetry
$SL(6,\R)$.


\chapter{Reduction from D=5 Gauged Supergravty}

The ungauged $N=8,D=5$ supergravity [\Cremmer]
has an $E_6$ rigid duality symmetry and 27 abelian vector fields
transforming as a {\bf 27} of $E_6$, which decomposes into
the $({\bf 15',1})+({\bf 6,2})$
under the subgroup   $SL(6,\R)\times SL(2,\R)$.
The gauge fields in the $({\bf 6,2})$ can be dualised to give massless
2-forms $B_{mn\, I\aa}$ and this is the form of the theory that was
gauged in [\gunwar,\PPV]. 
As before, $I,J=1,..,6$ are $SL(6,\R)$ indices, $\aa,\bb=1,2$ are $SL(2,\R)$
indices, while  $m,n=0,1,...,4$ are $D=5$ coordinate indices. In   [\gunwar],
gauged theories were constructed with    gauge groups $K=SO(p,6-p)$ for
$p=0,1,2,3$. In the gauged theory the subgroup 
$K=SO(p,6-p)$ of the rigid 
$SL(6,\R)$ symmetry is promoted to a local symmetry
in which the
vector fields $A_{m}^{ IJ}$ in the $({\bf 15',1})$ of $SL(6,\R)\times
SL(2,\R)$ become the non-abelian gauge fields transforming in the
adjoint of $K$.
In the gauged theory, the 2-forms
$B_{mn\, I\aa}$  become massive self-dual gauge fields
satisfying
a constraint of the form 
$$DB_{I\aa} =g  M_{I\aa}{}^{J\bb} *B_{J\bb}
\eqn\cons$$
for some    mass matrix
$M$, and each massive self-dual gauge field
 has three degrees of freedom,      the same
number   as a massless 2-form or
vector. The action contains the terms
$$
\int {1\over 2g} \eta ^{IJ} \ee ^{\aa\bb} B_{I\aa}\wedge DB_{J\bb}
+ 
{1\over 2} M^{I\aa J\bb}B_{I\aa}\wedge * B_{J\bb}
\eqn\acti$$
where $\eta _{IJ}$ is the constant $SO(p,6-p)$-invariant metric, $g$ is
the gauge coupling constant and $M^{I\aa J\bb}$ is a
scalar-dependent  mass matrix given explicitly in [\gunwar].
$D$ is the $SO(p,6-p)$ gauge covariant exterior derivative
$$ DB_{I\aa}=dB_{I\aa}-g \eta _{IJ}A^{JK}\wedge B_{K\aa}
\eqn\abc$$
Varying the action \acti\ gives the constraint \cons\ with
$M_{I\aa}{}^{J\bb}= \ee_{\aa\gg} \eta_{IK}
 M^{K\gg J\bb}$. The full action of [\gunwar] has an  additional coupling of 
the
form
$\int B\cdot J$ for a certain 2-form current $J$ constructed from the other 
fields
in the theory, giving   additional terms involving $J$ to \cons\ and some of 
the
equations below, but these will be suppressed here to simplify the 
presentation;
they do not affect the results. The action \acti\
 is invariant under local $SO(p,6-p)$ gauge transformations with
$B_{I\aa}$ transforming as a {\bf 6}.
As the 2-forms are massive, there is no 2-form gauge invariance, although
a form of the theory with such symmetry can be constructed by introducing
Stuckelberg 1-form gauge fields
$a_{I\aa}$ by the field redefinition
$$B_{I\aa} \to B_{I\aa} -Da_{I\aa}
\eqn\abc$$
which also introduces the gauge invariances
$$
\dd B_{I\aa}  =D\ll_{I\aa} ,\qq 
\dd a_{I\aa}  =\ll_{I\aa}
\eqn\abc$$
with 1-form parameters
$\ll_{I\aa}$.
Clearly, the 1-forms $ a_{I\aa} $ can be gauged to zero, regaining the previous
formulation. The 
  2-forms can be thought of as obtaining their mass by eating the 1-forms.
The minimal couplings break $SL(6,\R)$ to $K$ but $SL(2,\R)$ remains as a
global symmetry.

The dimensional reduction to $D=4$ of most sectors of the theory is
straightforward, but the
gauge and 2-form sectors have some unusual features.
 The reduction of the gauge fields gives vector gauge fields
$A_{\mm}^{ IJ}$ and scalars $\ss ^{ IJ}$ in the adjoint of
$K=SO(p,6-p)$  while that of the metric gives the $D=4$ metric, a
vector field $C_\mm$ and a scalar $\phi$.
The resulting theory clearly has a gauge symmetry which includes
$K\times U(1)$.
The
2-forms $B_{mn\, I\aa}$ give
2-forms $B_{\mm\nn\, I\aa}$
and vector fields 
 $A_{\mm \, I\aa}$
and the dimensional reduction of
the self-duality constraint
\cons\
now gives the constraint  
$$B_{I\aa} =g M_{I\aa}{}^{J\bb} *DA_{J\bb} 
\eqn\abc$$
so that the
2-forms $B_{\mm\nn\, I\aa}$ are dual to 
the vector fields 
 $A_{\mm \, I\aa}$
and the theory can be formulated in terms of  $A_{\mm \, I\aa}$ alone.
To see this in more detail, the
dimensional reduction of
 \acti\ includes the terms
$$\eqalign{
&\int {1\over g} \eta ^{IJ} \ee ^{\aa\bb} (B_{I\aa}\wedge DA_{J\bb}
-\2 dC\wedge A_{I\aa}\wedge A_{J\bb})
\cr
&
+ 
{1\over 2} M^{I\aa J\bb}\left[ (B_{I\aa}-C\wedge A_{I\aa})
\wedge * (B_{J\bb}-C\wedge A_{J\bb})
+ e^{-2\phi/\sqrt 3}
A_{I\aa}
\wedge *   A_{J\bb}\right]
\cr}
\eqn\abc$$
The field equation for $B$ gives $B$ in terms of the other fields, and
using this to eliminate $B$ from the action gives
$$\eqalign{
&\int 
\2 \ti M_{I\aa J\bb} F^{I\aa}\wedge * F^{J\bb}
- \eta _{IJ} \ee _{\aa\bb}  
 g dC\wedge A^{I\aa}\wedge A^{J\bb}
\cr
&
+ 
{1\over 2}g^2 e^{{-2\phi/\sqrt 3}}  \hat M_{I\aa J\bb}
A^{I\aa}
\wedge *   A^{J\bb}
\cr}
\eqn\actf$$
where
$$A^{I\aa} = g\eta ^{IJ} \ee ^{\aa\bb}  A_{J\bb},\qq
F^{I\aa}=DA^{I\aa}=dA^{I\aa}-g\eta _{JK}A^{IJ}\wedge A^{K\aa}
\eqn\abc$$
and
$$\ti M_{I\aa J\bb} = (M^{-1})_{I\aa J\bb},\qq
 \hat M _{I\aa J\bb} =\ee_{\aa\gg}\ee_{\bb\dd}\eta _{IK}\eta _{JL}
 M ^{K\gg L\dd}\eqn\abc$$

There is then local $SO(p,6-p)\times U(1)$ gauge symmetry
with gauge fields
$A^{IJ},C$ and there
remains a global $SL(2,\R)$ symmetry.
There are massive vector fields $A^{I\aa}$ transforming
as a $({\bf 6,2})$ under
$SO(p,6-p)\times SL(2,\R) $ and which are singlets under $U(1)$.
They transform non-trivially under the gauge group, and   their
field strength is $F=DA$; as these massive vector fields have no
gauge invariance in this formulation, there is no problem with this
coupling.  The massive vector fields can become gauge fields
by introducing a Stuckelberg scalars
via the field redefinition
$$A^{I\aa} \to A^{I\aa} -D \rho ^{I\aa} 
\eqn\abc$$
so that the theory now has   extra gauge invariances
$$\dd A^{I\aa} = D\ll ^{I\aa} , \qq \dd
\rho ^{I\aa} = \ll ^{I\aa} \eqn\stuck$$
where $D$ is the $K$-covariant derivative.
The gauge fields
$A^{IJ},A^{I\aa},C$ might then be thought of as gauge fields for the
gauge group
$(SO(p,6-p) \sd \R^{12}) \times U(1)$, but there remains the unusual
coupling $dC\wedge A\wedge A$ in \actf.
This coupling has an interesting topological interpretation [\topo].
The terms involving $C$   then include
$$\int 
 \2 e^{-\phi/\sqrt 3}dC\wedge *dC
- \eta _{IJ} \ee _{\aa\bb}  
 g dC\wedge A^{I\aa}\wedge A^{J\bb}
\eqn\gfgsf$$
where the kinetic term arises from the reduction of the Einstein action. 
As $C$ occurs only through its field strength,
it can be dualised to a new gauge field $\ti C$, so that \gfgsf\ becomes
$$\int 
 \2 e^{-\phi/\sqrt 3}G\wedge *G
\eqn\abc$$
where
$$G =d\ti C +  g \eta _{IJ} \ee _{\aa\bb}  
    A^{I\aa}\wedge A^{J\bb}
\eqn\abc$$
Comparing with
\fiss, one sees that $F^{IJ}, F^{I\aa},F^{\aa\bb}=G\ee^{\aa\bb}$ have the 
correct
form to be identified with the field strengths of the $CSO(p,6-p,2)$ gauging, 
and 
it is straightforward to check that the reduction and duality
transformation indeed give the
$CSO(p,6-p,2)$ gauged $D=4$ supergravity.
The scalars consist of the 42 scalars of the $D=5$ theory, taking values
in $E_6/USp(8)$, and the $15+1+ 12$ scalars
$\ss ^{ IJ},\phi,\rho ^{I\aa} $, giving $70 $ scalars in all, which is the
correct number
for $N=8,D=4$ supergravity.
The 12 scalars $\rho ^{I\aa} $ are eaten by the gauge fields $ A^{I\aa}$
which become massive, breaking the gauge group to
$K\times U(1)$.
As in the $D=5$ theory, the gauge group is further broken
to a subgroup of the   maximal compact subgroup $SO(p)\times
SO(6-p)\times U(1)$, with further vectors becoming massive through the
Higgs mechanism. 
The vacua correspond to critical points of the potential, and the
scalar expectation values will determine which subgroup of
$SO(p)\times
SO(6-p)\times U(1)$ remains unbroken.
 The $SL(2,\R)$ global symmetry remains, but is broken to $SL(2,\Z)$ in the
quantum theory.

Consider next the $SU(3,1)=SO^*(6)$ gauging of $D=5,N=8$ supergravity
[\GRW] and its reduction to $D=4$.
The structure of the gauging is  identical in form to that of the $SO(6)$
gauging.
Instead of focusing on the
$SL(6,\R)\times SL(2,\R)$ subgroup
of $E_6$, 
the starting point in [\GRW] was to consider the
  $SU^*(6 )\times SU(2)$ subgroup, under which 
the {\bf 27} decomposes as
the $({\bf {\overline {15} },1})+({\bf 6,2})$.
Again, it is useful to
introduce $SU^*(6 ) $ indices
$I',J'=1,..,6$ and $SU(2)$ indices   $\aa ',\bb '=1,2$.

The gauge fields in the $({\bf 6,2})$ can be dualised to give massless
2-forms $B_{mn\, I'\aa'}$ and this is the form of the theory that was
gauged in [\GRW]. 
  In the $SU(3,1)$ gauged theory [\GRW] the subgroup 
$K=SO^*(6)\sim SU(3,1)$ of the rigid 
$SU^*(6 ) $ symmetry is promoted to a local symmetry
in which the
vector fields $A_{m}^{ I'J'}$ in the $({\bf \overline {15} ,1})$   become
the non-abelian gauge fields transforming in the adjoint of $SO^*(6)$.
The   $B_{mn\, I'\aa'}$ become massive self-dual 2-forms
and the $SO^*(6)$ gauged theory can be written in the
$SU^*(6 )\times SU(2)$ basis so that it is identical in form to the
$SO(6)$ gauged theory   written in the
$SL(6,\R)\times SL(2,\R)$
 basis, but with indices $I,J$ and $\aa,\bb$ replaced by primed indices
$I',J'$ and $\aa',\bb'$.
In particular the
terms in the action involving the 2-forms include
$$
\int {1\over 2g} \eta ^{I'J'} \ee ^{\aa'\bb'} B_{I'\aa'}\wedge DB_{J'\bb'}
+ 
{1\over 2} M^{I'\aa 'J'\bb'}B_{I'\aa'}\wedge * B_{J'\bb'}
\eqn\abc$$
and the dimensional reduction to $D=4$ is identical in form to that
considered above
for the $SO(6)$ gauging. Then
dimensionally reducing the $SO^*(6)$ theory to $D=4$ and
   dualising the 
graviphoton $C$ as above gives precisely the
$CSO^*(6,2)$ gauged theory considered in section 4.
This has an $SU(2)$ global symmetry inherited from that of the 
$D=5$ theory.

\chapter{Acting with Dualitites}

In  the $CSO(p,q,r)$ gaugings, the $r(r-1)/2$ gauge fields
$A^{\aa\bb}$ only appear through their exterior derivative
$dA^{\aa\bb}$, and so any set of these can be dualised to
give a dual
form of the theory, and in general this changes the form of the gauge group.
This was seen explicitly in the last section for the
  $CSO(p,q,2)$ gauging, in which dualising 
$A^{\aa\bb}=\ti C\ee^{\aa\bb}$   to a gauge field $C$
gives the dual form with gauge group
$(SO(p,6-p) \sd \R^{12}) \times U(1)$
that arises from dimensional reduction of the $D=5$, $SO(p,q)$ gauge theory.
Similarly, dualising the
$CSO^*(6,2)$ theory gave the
$(SO^*(6) \sd \R^{12}) \times U(1)$ gauging that arises from the reduction
of the
$SO^*(6)$ gauging in $D=5$. Such dual forms give classically equivalent 
theories,
even though the form of the gauge group changes.

The $SO(p,q)$ and $CSO(p,q,r)$ gauged $N=8$ supergravities were obtained
from the $SO(8)$ gauging by acting with dualities
  [\nct,\nctt].
The    $SO(8)$ gauging is constructed from the ungauged theory
by adding minimal couplings and the couplings \vis,\ferms\
constructed from the T-tensor \tis.
These additional couplings all 
 involve the metric $\dd_{ab}$
and explicitly break the $SL(8,\R)$ symmetry of the ungauged action
to the subgroup $SO(8)$ preserving $\dd_{ab}$.
Acting on the gauged theory with an $SL(8,\R)$ transformation $S_a{}^b$
has the net effect of replacing every occurrence of  
$\dd_{ab}$ in the gauged theory with
$$ \eta _{ab}=
S_a{}^cS_b{}^d\dd_{cd}
\ek
This is of course simply a field redefinition, giving an equivalent
theory.
However, every occurrence of  $\dd_{ab}$ 
in the gauged theory is accompanied by a factor of the gauge coupling 
$g$. Choosing
$S=exp(-tX/2)$ where $X$ is the $SO(p)\times SO(q)$ invariant generator
$$X_{ab}= \pmatrix{
\aa {\bf 1}_{p \times p} & 0   \cr
0 & \bb {\bf 1}_{q \times q}  
}\eqn\abc$$
and 
$$\aa=-1, \qq \bb= p/q, \qq p+q=8
\ek
and rescaling $g \to ge^{\aa t}$ has the net effect of replacing
every occurrence of  $\dd_{ab}$ 
in the gauged theory with
$$ \eta_{ab} =\pmatrix{
{\bf 1}_{p \times p} & 0  \cr
0 & \xi {\bf 1}_{q \times q}  
}\eqn\abc$$
where $$\xi =e^{(\aa-\bb)t}\ek
Then this gives a well-defined theory for all real values of the
parameter $\xi$, and in particular $\xi $ can be continued to zero or to
negative values.
The limit $t\to \infty$, $\xi \to 0$ gives  the
$CSO(p,q)$ gauging, while continuing to 
$\xi =-1$, $t=i\pi/(\aa-\bb)$ gives the 
$SO(p,q)$ gauging.
The one-parameter family gives three distinct gaugings:
any theory with $\xi>0$ is 
equivalent (via field redefinitions) to the $SO(8)$ gauging, while
any theory with $\xi<0$ is 
equivalent  to the $SO(p,q)$ gauging. The power of this method is that
since 
supersymmetry is guaranteed for all $\xi >0$, it follows by  
continutation that the deformed theory will also be supersymmetric for
$\xi\le 0$;
 see [\nct,\nctt] for further details.

Similarly, a 2-parameter family of gaugings can be obtained [\nctt]
by acting with the $SL(8,\R)$ transformations 
$$S=exp(-tX/2-t'X'/2)
\ek
where
$$X_{ab}= \pmatrix{
\aa {\bf 1}_{p  \times p } & 0   \cr
0 & \bb {\bf 1}_{q+r \times q+r}  
},
\qq 
X'_{ab}=  \pmatrix{
\aa ' {\bf 1}_{p +q\times p+q} & 0   \cr
0 & \bb '{\bf 1}_{r \times r}  
}  
\eqn\abc$$
and 
$$\aa=\aa'=-1, \qq \bb=p/q+r , \qq \bb'=p+q/r,\qq p+q+r=8
\ek
and rescaling $g \to ge^{\aa t+\aa't'}$.
This has the effect of replacing every occurrence of
 $\dd_{ab}$ 
in the gauged theory with
$$ \eta_{ab} =\pmatrix{
{\bf 1}_{p \times p} & 0 &0 \cr
0 & \xi {\bf 1}_{q \times q} &0\cr 0& 0& \xi \zeta {\bf 1} _{r\times r}
}\eqn\abc$$     
where
$$\xi =e^{(\aa-\bb)t}, \qq \zeta =e^{(\aa '-\bb ')t'}\ek
This 2-parameter family  again divides into equivalence classes, and
contains 
the $CSO(p,q,r)$ gauging arising when
$\xi=-1, \zeta =0$.
For $\zeta >0$, the $\xi$-family consists of the
$SO(8),SO(p,q+r)$ and $CSO(p,q+r)$ gaugings, when
 $\zeta <0$, the $\xi$-family consists of the
$SO(p+q,r),SO(p+r,r)$ and $CSO(p,q+r)$ gaugings while
for
 $\zeta =0$, the $\xi$-family consists of the
$CSO(p+q,r),CSO(p,q,r)$ and $CSO(p,q+r)$ gaugings.

The same continuation techniques can be used to generate the
$CSO^*(2p,2q)$ gaugings from the
$SO^*(8)$ gauging.
Starting with the 
$SO(6,2)=SO^*(8)$ gauging and acting with
the $SU^*(8)$ transformation
$S=exp(-tX/2)$  
where $X$ is the $SO^*(2p)\times SO^*(2q)$ invariant generator
$$X_{a'b'}= \pmatrix{
\aa {\bf 1}_{2p \times 2p} & 0   \cr
0 & \bb {\bf 1}_{2q \times 2q}  
}\eqn\abc$$
and 
$$\aa=-1, \qq \bb= p/q, \qq p+q=4
\ek
and rescaling $g \to ge^{\aa t}$ has the net effect of replacing
   $\dd_{ab}$ 
in the gauged theory with
$$ \eta_{ab} =\pmatrix{
{\bf 1}_{2p \times 2p} & 0  \cr
0 & \xi {\bf 1}_{2q \times 2q}  
}\eqn\abc$$
where $$\xi =e^{(\aa-\bb)t}\ek
The limit $t\to \infty$, $\xi \to 0$ gives the $CSO^*(2p,2q)$
gauging, but continuing to $\xi=-1$ recovers one of the gaugings already
considered and doesn't give anything new.
The fact that the $CSO^*(2p,2q)$ gauging can be obtained as a smooth
limit of a 1-parameter family of consistent supersymmetric theories
guarantees that the
$CSO^*(2p,2q)$ gauged theory exists and is supersymmetric, by the
arguments of [\nct].

An alternative way of obtaining the 
$CSO^*(6,2)$ theory is as follows.
It was seen in section 5 that dimensionally reducing the 
$D=5$, $SO(6)$ gauged theory gives a $D=4$ theory with gauge group
$(SO(6)\sd \R^{12}) \times U(1)$, which can be dualised to 
the
$CSO(6,2)$ theory by dualising the $U(1)$ gauge field, while
dimensionally reducing the 
$D=5$, $SO^*(6)$ gauged theory gives a $D=4$ theory with gauge group
$(SO^*(6)\sd \R^{12}) \times U(1)$, which can be dualised to 
the
$CSO^*(6,2)$ theory by dualising the $U(1)$ gauge field.
Before dualising the 
$U(1)$ gauge field, the theory with
gauge group $(SO(6)\sd \R^{12}) \times U(1)$ and the one with
gauge group 
$(SO^*(6)\sd \R^{12}) \times U(1)$ are both deformations of the
form of the $D=4,N=8$ theory with
global symmetry of the action $L=E_6\times \R$, which is the dual form
obtained directly  by reducing the ungauged $D=5$ theory.
They can be thought of as gaugings of different subgroups of $E_6\times \R^+$.
This means that one can act on the
the theory with
gauge group $(SO(6)\sd \R^{12}) \times U(1)$
using an $E_6$ transformation to change the couplings.
In the $SL(6,\R)\times SL(2,\R)$ basis,
the generators of $E_6$ consist of
$SL(6,\R)$ and $ SL(2,\R)$
generators, together
with
generators
$\sss_{IJK\aa}$ in the $({\bf 20',2})$
of $SL(6,\R)\times SL(2,\R)$ satisfying 
$$\sss_{IJK\aa}=\sss_{[IJK]\aa}= \6 \ee_{IJKLKM} \ee_{\aa\bb}\sss^{LKM\bb}
\ek
where $\sss^{IJK\aa}\equiv (\sss_{IJK\aa})^*$.
As in [\GRW,\gunwar], let
$$X_{abcd}=- 
\dd_{abcd}^{1357}
+\dd_{abcd}^{2468}
+\dd_{abcd}^{1368}
-\dd_{abcd}^{2457}
+\dd_{abcd}^{1458}
-\dd_{abcd}^{2367}
+\dd_{abcd}^{1467}
-\dd_{abcd}^{2358}
\eqn\xiss$$
and decompose $a,b=1,...,8$ into
indices $I,J=1,...,6$ and $\aa,\bb=7,8$ to define
a generator
$X_{IJK\aa}$ of $E_6$.
Then acting with the $E_6$ transformation $\exp(tX)$, rescaling the
coupling constants as above and taking the limit $t\to i\pi/4$ 
takes the
 theory with
gauge group $(SO(6)\sd \R^{12}) \times U(1)$ to the one with
gauge group 
$(SO^*(6)\sd \R^{12}) \times U(1)$.

\chapter{The Scalar Potentials}

The scalar potentials of the $D=4$, $SO(8)$ gauging was analysed in [\schur], 
and
those of the non-compact $D=4$ gaugings in [\HWpot].
The theories with gauge groups
$SO(4,4)
$ and
$SO(5,3)$ have de Sitter vacua arising at local maxima of the potentials 
[\nctt,\DS].  The $CSO(2,0,6)$ gauging has a Minkowski space solution
and the potential has flat directions [\nctt].
The structure and potentials of these models were analysed further in
[\HWpot]; no other critical points are known.
All of these therories   have domain wall solutions that preserve half the
supersymmetry [\wall].

In section 5 it was seen that the dimensional reduction of the $N=8,D=5$
theories with gauge groups
$SO(p,6-p)$ or $SO^*(6)$ gives (after dualising the graviphoton) the 
 $N=8,D=5$
theories with gauge groups
$CSO(p,6-p,2)$ or $CSO^*(6,2)$.
If $V_5(\chi)$ is the $D=5$ scalar potential depending on 42 scalars
$\chi$ in $E_6/USp(8)$, then its dimensional reduction gives
a scalar potential
$$V_4(\chi,\phi)= e^{2\phi/\sqrt 3 }V_5(\chi)
\eqn\abc$$
where $\phi$ is the 
scalar coming from the reduction of the metric.
The full scalar potential $V(\chi,\phi,\ss,\rho)$ of the $D=4$ theories
also depends in principle on the scalars
$\ss ^{IJ}$ or $\ss ^{I'J'}$ from the reduction of the $D=5$ gauge fields
and on the Stuckelberg scalar fields $\rho^{I\aa}$ or
$\rho^{I'\aa'}$, but in fact is independent of the
Stuckelberg scalars $\rho$  as the potential is necessarily invariant under the
gauge invariance \stuck.
Thus the potentials have (at least) 12 flat directions
and in addition are invariant under the global symmetry $SL(2,\R)$ or
$SU(2)$ as well as the local $SO(p,6-p)\times USp(8)$ or
$SO^*(6)\times USp(8)$
symmetries. 
Note that the exponential dependence on $\phi $ implies that
$V_4$ can only have critical points
at  values $\chi=\chi_0$ which are critical points of $V_5$,
$\partial V_5(\chi_0)=0$, at which
the potential vanishes, $V_5(\chi_0)=0$.
 However, the full 
potential $V$   has terms with different $\phi$ dependence.
For example, dimensional reduction of the $D=5$ vector kinetic term $N_{IJ\,
KL}(\chi) F^{IJ}\cdot F^{KL}$ for
the vector fields $A^{IJ}$
gives a contribution to the scalar potential of the form
$$e^{ {-2\sqrt 3\phi }}N_{IJ\, KL}(\chi) \ss^I\ss^J \ss^K\ss^L
\eqn\abc$$
that is quartic in $\ss$, where $N_{IJ\, KL}(\chi)=N_{[IJ][KL]}(\chi)$
is a function of the $D=5$ scalars $\chi$.
The full  $D=4$ potential 
could then in principle have critical points even if the $D=5$ potential $V_5$ 
or
the restricted $D=4$ potential $V_4$    do not.

The structure of the $D=5$ potentials $V_5$ was analysed in
[\gunwar,\GRW], and this can now be used to obtain information about the
$D=4$ potentials.
For the $SO(p,6-p)$ gaugings it is straigthforward to find the dependence
on scalars in  the
$SL(6,\R)/SO(6) \times SL(2,\R)/SO(2)$ subspace of
$E_6/USp(8)$, parameterised by a matrix $S_I{}^J$ in $SL(6,\R)$ and a
matrix
$S'_\aa{}^\bb$  in $SL(2,\R)$.
From [\gunwar], one finds
$$V_4= -{g^2\over 32}e^{2\phi/\sqrt 3 }\left[ \{ \tr (\eta M) \} ^2
-2 \{ \tr (\eta M \eta M) \} \right]
\eqn\abc$$
where $M_{IJ}$ is the symmetric matrix
$$M_{IJ}=S_I{}^KS_J{}^L\dd_{KL}
\eqn\abc$$
and the trace $\tr$ is taken over the six $SL(6,\R)$ indices.
This is independent of 
$S'$, as expected from $SL(2,\R)$ invariance.

The explicit dependence  of the potential $V_5$
on an $SO(p)\times SO(6-p)$ invariant scalar $\ll$ 
 was found in [\gunwar].
By the Schur's lemma argument of [\schur], a critical point of the potential 
$V_5$
restricted to such an $SO(p)\times SO(6-p)$ invariant scalar will in fact
be a critical point of the full potential.
 Let
$$\eta_{IJ}= diag (1_p, \xi 1_q)
\eqn\abc$$
where $\xi=1 $ for the $SO(6)$ gauging and 
$\xi=-1 $ for the $SO(p,6-p)$ gauging.
For the $SO(5)$ invariant direction, taking
$$M=diag (e^\ll,...,e^\ll,e^{-5\ll})
\eqn\abc$$
gives 
$$V_5 = -{g^2\over 32}  \left[
15e^{2\ll}+10 \xi e^{-4\ll}-e^{-10\ll} \right] 
\eqn\vfi$$
For the $SO(4)\times SO(2)$ invariant direction, taking
$$M=diag (e^\ll,...,e^\ll,e^{-2\ll},e^{-2\ll})
\eqn\abc$$
gives 
$$V_5 =-{g^2\over 4}   \left[
e^{2\ll} +2\xi e^{-\ll} \right] 
\eqn\vfo$$
For the scalars in the $SO(3)\times SO(3)$ invariant direction contained
in $SL(6,\R)$, taking
$$M=diag (e^\ll,e^\ll,e^\ll,e^{-\ll},e^{-\ll},e^{-\ll})
\eqn\vth$$
gives 
$$V_5=-{3g^2\over 16}   \left[
\cosh {(2\ll)} +3\xi  \right] 
\eqn\abc$$

The $D=5$, $SO(6)$ gauging has a maximally supersymmetric AdS critical
point at which all scalars vanish (corresponding to $\ll=0,\xi=1$ in each
of the above cases \vfi,\vfo,\vth)
giving a $D=4$ potential which, on setting all scalars but $\phi$ to
zero, is
$V_4 =-3g^2e^{2\phi/\sqrt 3 }/4
$.
The $D=5$, $SO(6)$ gauging also has a non-supersymmetric  
$SO(5)$-invariant AdS  critical point with $e^{6\ll}=1,\xi=1$ in \vfi.
The $D=5$, $SO(3,3)$ gauging has a maximally supersymmetric de Sitter
critical point at which all scalars vanish (corresponding to
$\ll=0,\xi=-1$ in \vth) giving 
a $D=4$ potential which, on setting all scalars but $\phi$ to
zero, is
$$V ={3\over 8}g^2e^{2\phi/\sqrt 3 }
\eqn\abc$$
This is precisely the form of potential 
that is required for quintessence, and
precisely this theory was proposed in [\quin] as one that
could give a cosmology with quintessence.
However, reinstating the scalar field $\ll$, the form of the potential
is
$$V_4={3g^2\over 16} e^{2\phi/\sqrt 3 }  \left[3-
\cosh {(2\ll)}    \right] 
\eqn\abc$$
This has the desired exponential roll in the $\phi$
direction, but is unbounded below in the $\ll$ direction and so it seems
  that the slow-roll solution used in [\quin] will be unstable to
fluctuations in the $\ll$ direction.

The $D=5,SO^*(6)$ gauged theory and its reduction can be treated in the
same way,  and 
  it is   straightforward to find the dependence
on scalars in  the
$SU^*(6)/USp(6) \times SU(2)/SO(2)$ subspace of
$E_6/USp(8)$, parameterised by a matrix $S_{I'}{}^{J'}$ in
$SU^*(6)$ and a matrix
$S'_{\aa'}{}^{\bb '}$  in $SU(2)$.
Remarkably, the potential is completely independent of 
both
$S_{I'}{}^{J'}$   and  
$S'_{\aa'}{}^{\bb '}$
 [\GRW], so that the potential has a large number of flat directions.
However, the potential is non-trivial in other directions, and we now turn to
its dependence on one of these other directions. 

The 42 scalar fields in $E_6/USp(8)$ can be parameterised by a
27-bein ${\cal V}$ taking values in $E_6$.
Choosing the ansatz ${\cal V}=\exp (\ll X)$, where $X$ is the $E_6$ generator
constructed from the
$X_{IJK\aa}$ defined in \xiss, picks out  a particular scalar field  $\ll(x)$ 
in
an
$SU(3)$-invariant direction
 [\gunwar,\GRW].
The dependence on $\ll$ of the  
potential of the $D=5,SO(6)$ gauging was calculated in [\gunwar] and for the
$D=5,SO^*(6)$ gauging was calculated in [\GRW]. These then give
the restricted $D=5$ potentials
$$V_5 (\ll)= {3g^2\over 32}  \left[
p^2-4\xi p-5 \right] 
\eqn\vfif$$
where
$$p=\cosh{(4\ll)}
\eqn\abc$$
  where the case $\xi=1$ is for the $SO(6)$
gauging
and the case $\xi=-1$ is for the $SO^*(6)$
gauging.
The restricted $SO(6)$ potential has the maximally supersymmetric critical
point  at
$p=1,\ll=0$ and a non-supersymmetric AdS critical point at
$p=2$, breaking the $SO(6)$ gauge symmetry to $SU(3)\times U(1)$
[\gunwar].

For the $SO^*(6)$ gauging, there is a critical point at $\ll=0$
at which the potential \vfif\ (with $\xi =-1$) 
vanishes.
This is then
a critical point of the full potential and the  $D=5$, $SO^*(6)$ gauging
has
a Minkowski space vacuum [\GRW].
It was shown in [\GRW] that this preserves $N=2$ supersymmetry,
breaks the gauge group to $SU(3)\times U(1)$. The potential $V_5$ is 
independent
of all the scalar fields  in the
 $SU^*(6)/USp(6) \times SU(2)/SO(2)$ subspace of
$E_6/USp(8)$
and so it has at least  16 flat directions.

As $V_5(0)=0$, the  critical point at $\ll=0$ is also a critical point
of $V_4$, and this is in fact a critical point for the full scalar potential.
 The $D=4,CSO^*(6,2)$   gauged theory then has a $D=4$ Minkowski-space
solution  preserving $N=2$
supersymmetry   and $SU(3)\times U(1)\times U(1)$ gauge symmetry, as well as
preserving the $SU(2)$ global symmetry.
There are at least 28 flat directions, corresponding to the
16 flat directions in $SU^*(6)/USp(6) \times SU(2)/SO(2)$ and the 12 
Stuckelberg
scalars $\rho$.


\refout

\bye